\documentstyle[lprocl,epsf]{article}
\catcode`\@=11
\def\mref#1{\ifx\und@fined#1
{need to supply reference \string#1.}
\else #1 \fi}
\catcode`\@=12 %
\let\lref=\def
\newif\ifhypertex
\ifx\hyperdef\UnDeFiNeD
    \hypertexfalse
    \def\hyperdef#1#2#3#4{#4}
    \def\hfoot#1{}%
    
    \def\e@tf@ur#1{}
    \def\eprt#1{{\tt #1}}
    \def\CERN{Theoretical Physics Division, CERN, Geneva, Switzerland}
    \def\WL{W.\ Lerche}
    
    \def\lp{XVIII International Symposium on
           Lepton Photon Interactions}
      
\else
    \hypertextrue
   \def\eprt#1{
  {\tt #1}}
\def\CERN{
   {Theoretical Physics Division,} 
  
 CERN, Geneva, Switzerland}
\def\hfoot#1{\foot{#1}}%
\def\WL{

 W.\ Lerche}
    
\def\lp{{{LP
'97}}}
   
\fi
\newcommand\inbar{\vrule height1.5ex width.4pt depth0pt}
\newcommand\IC{\relax\,\hbox{$\inbar\kern-.3em{\rm C}$}}
\newcommand\IQ{\relax\,\hbox{$\inbar\kern-.3em{\rm Q}$}}
\newcommand\IR{\relax{\rm I\kern-.18em R}}
\newcommand\IP{\relax{\rm I\kern-.18em P}}
\newcommand\ZZ{\relax{\hbox{\rm Z\kern-.42em Z}}}

\newcommand\abs{
After briefly reviewing basic concepts of perturbative string theory,
we explain in simple terms some of the new findings that created
excitement among the string physicists. These developments include
non-perturbative dualities and a unified picture that embraces the
so-far known theories.
}

\newcommand\proc{{{\large
{\it $^\dagger$Contribution to the Proceedings of the
XVIII International
Symposium on Lepton Photon
Interactions {\lp}, Hamburg, July 1997.}}
This overview for non-experts is not meant to contain exhaustive
references.} }

\begin{document}
\begin{titlepage}
\topskip0.5cm
\hfill\hbox{CERN-TH/97-299}\\[-.3cm]
\flushright{\hfill\hbox{hep-th/9710246}}\\[3.3cm]

\begin{center}{\Large\bf
{Recent Developments in String Theory$^\dagger$}\\[2.5cm]}{
\large \WL\\[.7cm]}
{\large \CERN}\\[3.cm]
\center{\large{\bf Abstract}}\\[.3cm]
\flushleft{\large{\abs}}
\end{center}
\vskip2.5cm
\flushleft{\large {\proc}}
\vfill
\hbox{CERN-TH/97-299}\hfill\\[-.4cm]
\hbox{October 1997}\hfill\\
\end{titlepage}
%



\title{RECENT DEVELOPMENTS IN STRING THEORY}

\author{W.LERCHE}

\address{\CERN}

\maketitle\abstracts\abs

\section{Perturbative String Theory}
\subsection{Basic features}\label{subsec:basic}

Heuristically one may introduce string theory\cite{intro}\ by
describing tiny loops sweeping through $D$-dimensional
space-time.  Being one-dimensional extended objects, they trace out
two-dimensional ``world-sheets'' $\Sigma$ that may be viewed as
thickened Feynman diagrams:

\epsfysize=.8in
{\centerline{\epsffile{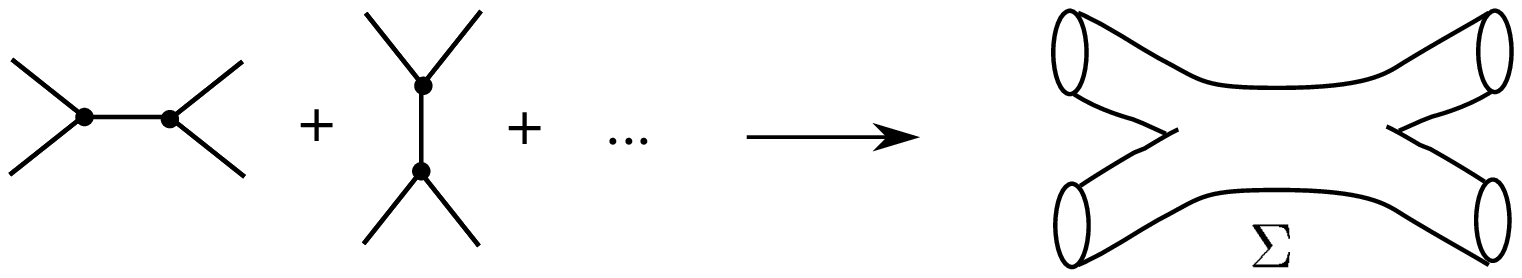}}}

Accordingly, perturbative string theories are defined by
two-dimensional field theories living on such Riemann surfaces.
Suitable two-dimensional field theories can be made out of a great
variety of building blocks, the simplest possibilities being given by
free two-dimensional bosons and fermions, eg., $X_i(z)$, $\psi_i(z)$.

By combining the two-dimensional building blocks according to certain
rules, an infinite variety of operators that describe \/{\em
space-time} fields can be constructed. Typically, the more complex the
operators become that one builds, the more massive the associated
space-time fields become. In total one obtains a finite number of
massless fields, plus an infinite tower of string excitations with
arbitrary high masses and spins. Schematically, among the most generic
massless operators are:
\begin{equation}
\begin{array}{rcl}
{\rm Dilaton\ scalar} &
\phi &
\eta^{\mu\nu}\bar\partial X_\mu\partial X_\nu\qquad (\mu,\nu=1,...,D)
\\
{\rm Graviton} &
g_{\mu\nu} &
\bar\partial X_\mu\partial X_\nu
\\
{\rm Gauge\ field} &
A_{\mu}^a &
\bar\partial X_\mu\partial X^a
\\
{\rm Higgs\ field} &
\Phi^{ab} &
\bar\partial X^a\partial X^b
\end{array}
\end{equation}
We see here how simple combinatorics provide an intrinsic and
profound unification of particles and their interactions: from the 2d
point of view, a gauge field operator is very similar to a graviton
operator, whereas the $D$-dimensional space-time properties of these
operators are drastically different. An important point is that
gravitons necessarily appear, which means that this kind of string
theories {\em imply} gravity. This is one of the main motivations for
studying string theory. Indeed, it is believed by many physicists
that string theory provides the only way to formulate a consistent
quantum theory of gravity.

Fermions (``electrons'' and ``quarks'') are obtained by changing the
boundary conditions of the two dimensional fermions $\psi$ on $\Sigma$.
These can be anti-periodic or periodic (the fields change sign or not
when transported around $\Sigma$), giving rise to fermionic or bosonic
fields in space-time, respectively. It turns out that whenever the
theory contains space-time fermions, it must have two dimensional
supersymmetry, and this is why such string theories are called {\em
superstrings}. It is however not so that the $D$-dimensional space-time
theory must be supersymmetric, although practically all models studied
so far do have supersymmetry to some degree. But this is to simplify
matters and keep perturbation theory under better control,\footnote{An
often-cited argument for low-energy supersymmetry is that perturbation
theory of non-supersymmetric strings tends to be unstable due to vacuum
tadpoles. However, even when starting with a supersymmetric theory,
this problem will eventually appear, namely when supersymmetry is
(e.g., at the weak scale) spontaneously broken. We therefore have at
any rate to find mechanisms to stabilize the perturbation series, which
means that vacuum tadpoles are no strong reason for beginning with a
supersymmetric theory in the first place. See especially
refs.\cite{GM,KD}\ for a vision how strings could be more clever and
avoid supersymmetry (similar ideas may apply to the ``hierarchy
problem'' as well); for a different line of thoughts, see
ref.\cite{cosmo}.} much like the study of supersymmetric Yang-Mills
theory makes the analysis much easier as compared to non-supersymmetric
gauge theories. String theory does not intrinsically predict space-time
supersymmetry, although it arises quite naturally.

That very simple, even free two dimensional field theories generate
highly non-trivial effective dynamics in $D$-dimensional space-time,
can be visualized by looking to the perturbative effective action:

\epsfysize=2.5in
\centerline{\epsffile{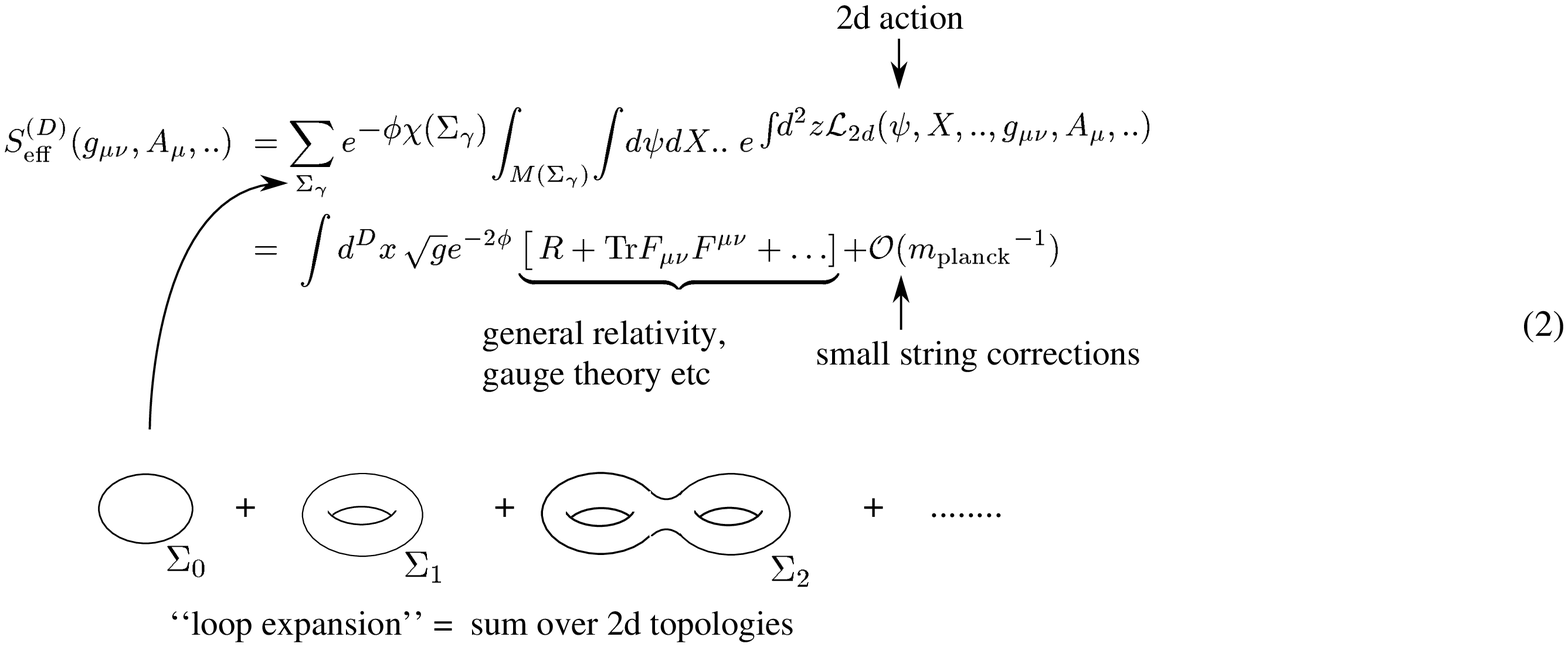}}

Here, $g_{\mu\nu},A_\mu,...$ are space-time fields that provide the
background in which the strings move, and ${\cal
L}_{2d}(\psi,X,..,g_{\mu\nu},A_\mu,..)$ is the lagrangian containing
the two-dimensional fields, as well as the background fields which
are simply coupling constants from the two dimensional point of view.
The 2d fields are integrated out, and one also sums over all the
possible two-dimensional world-sheets $\Sigma_\gamma$ (as well as
over the boundary conditions of the $\psi(z)$). This corresponds to
the well-known loop expansion of particle QFT. Note, however, that
there is only {\rm one} ``diagram'' at any given order in string
perturbation theory.

The topological sum is weighted by $e^{-\phi\chi(\Sigma_\gamma)}$,
where $\phi$ is the dilaton field and where the Euler number
$\chi(\Sigma_\gamma)\equiv 2-2\gamma$ is the analog of the
loop-counting parameter. The coupling constant for the perturbation
series thus is
$$
g=e^{\langle\phi\rangle},
$$
which must be small in order for the perturbation series to make
sense. We see here that a coupling constant is given by an {\em a
priori} undetermined vacuum expectation value of some field, and this
reflects a general principle of string theory.

In addition, the topological sum is augmented by integrals over the
inequivalent shapes that the Riemann surfaces can have, and this
corresponds to the momentum integrations in ordinary QFT. There may be
divergences arising from degenerate shapes, but these divergences can
always be interpreted as IR divergences in the space-time sense. In
particular, the well-known logarithmic running of gauge couplings in
four dimensional string theories arises from such IR divergences.

The absence of genuine UV divergences was another early motivation of
string theory, and especially makes consistent graviton scattering
possible: remember that ordinary gravity is not renormalizable and one
cannot easily make sense out of graviton loop diagrams. A very
important point is the origin of this well-behavedness of the string
diagrams. It rests on {\em discrete reparametrizations} of the
string-world sheets $\Sigma_\gamma$ (``modular invariance''), which
have no analog in particle theory. The string ``Feynman rules'' are
very different, and cannot even be approximated by particle QFT. String
theory is therefore {\em more} than simply combining infinitely many
particle fields, and it is this what makes a crucial difference. The
whole construction is very tightly constrained: modular invariance
determines the whole massive spectrum, and taking any single state away
from the infinite tower of states would immediately ruin the
consistency of the theory.

So far, we described the ingredients that go into computing the
perturbative effective action in $D$-dimensional space-time. Now we
focus on the outcome. As indicated in (2), the effective action
typically contains (besides matter fields) general relativity and
non-abelian gauge theory, plus stringy corrections thereof. These
corrections are very strongly suppressed by inverse powers of the
Planck mass, $m_{{\rm Planck}}\sim 10^{19}$GeV, which is the
characteristic scale in the theory. The value of this scale is
dictated by the strength of the gravitational coupling constant, and
governs the size or tension of the strings, as well as the level
spacing of the excited states. That strings are so difficult to
observe, and characteristic corrections so hard to see, stems from
the fact that the gravitational coupling is so small, and this is not
at the string physicists' fault -- they have no option to change that !

That general relativity, with all its complexity just pops
out from the air (arising from eg., a free 2d field theory), may sound
like a miracle. Of course this does not happen by accident, but is
bound to come out. The important point here is that there is a special
property that the relevant two dimensional theories must have for
consistency, and this is {\em conformal invariance}. This is a quite
powerful symmetry principle, which guarantees, via Ward identities,
general coordinate and gauge invariance in space-time -- however, only
so if and only if  $D=10$.

As we will see momentarily, this does not imply that superstrings
must live exclusively in ten dimensions, but it means that
superstrings are most naturally formulated in ten dimensions.

\subsection{10d Superstrings and their compactifications}

Two-dimensional field theories can have two logically independent,
namely holomorphic and anti-holomorphic pieces. For obtaining $D=10$
string theories, each of these pieces can either be of type superstring
``$S$'' or of type bosonic string ``$B$'' (with extra $E_8\times E_8$
or $SO(32)$ gauge symmetry). By combining these building blocks in
various ways,\footnote {There are further, but non-supersymmetric
theories in $D=10$.} plus including an additional ``open'' string
variant, one obtains the following exhaustive list of supersymmetric
strings in $D=10$:$\,$
\begin{center}
\begin{tabular}{|c c c l|}
\hline
Composition & Name &  Gauge Group & Supersymmetry\\
\hline
$S\, \otimes \bar S^\dagger$ & Type IIA & \ \ $U(1)$ & non-chiral
$N=2$ \\ $S\, \otimes \bar S$ & Type IIB & \ \ - & chiral $N=2$ \\
$S\, \otimes \bar B$ & Heterotic & $E_8\times E_8$ & chiral $N=1$ \\
$S\, \otimes \bar B'$ & Heterotic' & $SO(32)$ & chiral $N=1$\\ $(S\,
\otimes \bar S)/Z_2$ & Type I (open) & $SO(32)$ & chiral $N=1$ \\
\hline
\end{tabular}
\end{center}

Since these theories are defined in terms of 2d world-sheet degrees of
freedom, which is intrinsically a perturbative concept in terms of 10d
space-time physics, all we can really say is that there are five
theories
in ten dimensions that are different in {\em perturbation theory}.
Their perturbative spectra are indeed completely different, their
number of supersymmetries varies, and also the gauge symmetries are
mostly quite different.

Now, we would be more than glad if strings would remain in lower
dimensions as simple as they are in $D=10$. Especially string theories
in $D=4$ turn out to be much more complicated. Specifically, the
simplest way to get down to four dimensions is to assume/postulate that
the space-time manifold is not simply $\IR^{10}$, but $\IR^4\times
X_6$, where $X_6$ is some compact six-dimensional manifold. If it is
small enough, then the theory looks at low energies, ie., at distances
much larger than the size of $X_6$, effectively four-dimensional. This
is like looking at a garden hose from a distance, where it looks
one-dimensional, while upon closer inspection it turns out to be a
three dimensional object.

The important good feature is that such kind of ``compactified''
theories are  at low energies exactly of the type as the standard model
of particle physics (or some supersymmetric variant thereof). That is,
generically such theories will involve non-abelian gauge fields, {\em
chiral} fermions and Higgs fields besides gravitation, and all of this
coupled together in a (presumably) consistent and truly unified
manner~! That theories that are able to do this have been found at all,
is certainly reason for excitement.

But apart from this generic prediction, more detailed predictions, like
the precise gauge groups or the massless matter spectrum, cannot be
made at present -- this is the dark side of the story. The reason is
that the compactification mechanism makes the theories {\em enormously}
more complicated and rich than they originally were in $D=10$. This is
intrinsically tied to properties of the possible compactification
manifolds $X_6$:

\noindent {\bf i)} There is a large number of choices for the
compactification manifold. If we want to have $N=1$ supersymmetry in
$D=4$ from e.g.\ a heterotic string, then $X_6$ must be a
``Calabi-Yau'' space\cite{brian}, and the number of such spaces is
perhaps 10$^4$. On top of that, one has to specify certain instanton
configurations, which multiplies this number by a very large factor.
{\em A priori}, each of these spaces (together with a choice of
instanton configuration) leads to a different perturbative matter and
gauge spectrum in four dimensions, and thus gives rise to a different
four dimensional string theory.

\noindent {\bf ii)}
Each of these manifolds by itself has typically a large number of
parameters; for a given Calabi-Yau space, the number of parameters can
easily approach the order of several hundred.

These parameters, or ``moduli'' determine the shape of $X_6$, and
correspond physically to vacuum expectation values of scalar
(``Higgs'') fields, similar to the string coupling $g$ discussed above.
Changing these VEVs changes physical observables at low energies, like
mass parameters, Yukawa couplings and so on. They enter in the
effective lagrangian as free parameters, and are not determined by any
fundamental principles, at least as far is known in perturbation
theory. Their values are determined by the choice of vacuum, much like
the spontaneously chosen direction of the magnetization vector in a
ferromagnet (that is not determined by any fundamental law either). The
hope is that after breaking supersymmetry, the continuous vacuum
degeneracy would be lifted by quantum corrections (which is
typical for non-supersymmetric theories), so that ultimately  there
would be fewer accessible vacua.

\begin{figure} \epsfysize=1.8in \centerline{\epsffile{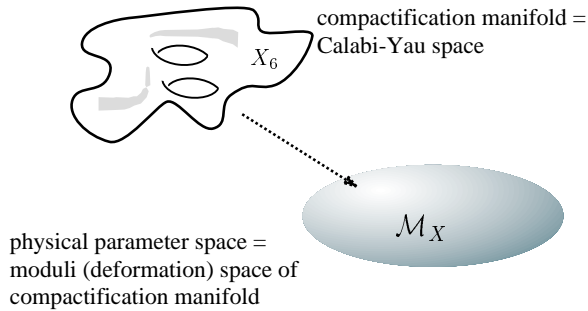}}
\caption{The parameter space of the low energy effective field theory
of a string compactification is essentially given by the
moduli space (deformation space) of the compactification manifold.
This leads to a geometric interpretation of almost all of the
parameters and couplings. Shown is that with each point in the moduli
space ${\cal M}_X$, one associates a compactification manifold $X$ of
specific shape; the shape in turn determines particle masses, Yukawa
couplings etc. The picture will be refined below.
\hfill
\label{fig:CYmod}}
\end{figure}

The situation is actually worse than described so far,
because we have on top of points i) and ii):

{\bf iii)} There are five theories in $D=10$ and not just one, and a
priori each one yields a different theory in four dimensions for a
given compactification manifold $X_6$. If one of them would be the
fundamental theory, what is then the r\^ole of the others ?

{\bf iv)} There is no known reason why a ten dimensional theory wants
at all to compactify  down to $D=4$; many choices of space-time
background vacua of the form $\IR^{10-n}\times X_n$ appear to be on
equal footing.

All these points together form what is called the {\em vacuum
degeneracy} problem of string theory -- indeed a very formidable
problem, known since a decade or so. Summarizing, most of the physics
that is observable at low energies seems to be governed by the vacuum
structure and not by the microscopic theory, at least as far as we can
see today. Still, it is not so that string theory would not make any
low-energy predictions;  the theories are very finely tuned and
internal consistency still dramatically reduces the number of possible
low energy spectra and independent couplings,  as compared to ordinary
field theory.

The recent progress in non-perturbative string theory does not solve
the problem of the choice of vacuum state either. The progress is
rather of conceptual nature and opens up completely new perspectives on
the very nature of string theory.

\section{Duality and non-perturbative equivalences}

Duality is the main new concept that has been stimulating the recent
advances in supersymmetric particle\cite{gaugerev} and string theory.
Roughly speaking, duality is a map between solitonic (non-perturbative,
non-local, ``magnetic'') degrees of freedom, and elementary
(perturbative, local, ``electric'') degrees of freedom. Typically,
duality transformations exchange weak and strong-coupling physics and
act on coupling constants like $g\rightarrow 1/g$. They are thus
intrinsically of quantum nature.

Duality symmetries are most manifest in supersymmetric theories,
because in such theories perturbative loop corrections tend to be
suppressed, due to cancellations between bosonic and fermionic degrees
of freedom. Otherwise, observable quantities get so much
polluted by radiative corrections that the more interesting
non-perturbative features cannot be easily made out.

More precisely, certain quantities (eg., Yukawa couplings) in a
supersymmetric low energy effective action are protected by
non-renormalization theorems, and those quantities are
typically {\em holomorphic} functions of the massless fields. As a
consequence, this allows to apply methods of complex analysis,
and ultimately of algebraic geometry, to analyze the physical theory.
Such methods (where they can be applied) turn out to be much more
powerful then traditional techniques of quantum field theory, and
this was the technical key to the recent developments.

A typical consequence of the holomorphic structure is a continuous
vacuum degeneracy, arising from flat directions in the effective
potential. The non-renormalization properties then guarantee that this
vacuum degeneracy is not lifted by quantum corrections, so that
supersymmetric theories often have continuous {\rm quantum moduli
spaces} $\cal M$ of vacua. In string theory, as mentioned above, these
parameter spaces can be understood geometrically as moduli spaces of
compactification manifolds.

\subsection{A gauge theory example}

One of the milestones in the past few years was the solution of the
(low energy limit of) $N=2$ supersymmetric gauge theory in four
dimensions.\cite{SW,SWreviews} Important insights that go beyond
conventional particle field theory have been gained by studying this
model, and this is why we briefly touch it here. In fact, even though
this model is an ordinary gauge theory, the techniques stem from string
theory, which demonstrates that string ideas can be useful also for
ordinary field theory.

Without going too much into the details, simply note that $N=2$
supersymmetric gauge theory (here for gauge group $G=SU(2)$) has a
moduli space $\cal M$ that is spanned by roughly the vacuum
expectation value of a complex Higgs field $\phi$. The relevant
holomorphic quantity is the effective, field dependent gauge coupling
$g(\phi)$ (made complex by including the theta-angle in its
definition). Each point in the moduli space corresponds to a
particular choice of the vacuum state. Moving around in $\cal M$ will
change many properties of the theory, like the value of the effective
gauge coupling $g(\phi)$ or the mass spectrum of the physical states.
An important aspect is that there are special regions in the moduli
space, where the theory behaves specially, ie., becomes {\em
singular}. This is depicted in Fig.2.

More precisely, there are two different types of such singular regions.
Near $\langle\phi\rangle\rightarrow \infty$, the gauge theory is weakly
coupled since the effective gauge coupling becomes arbitrarily small:
$g(\phi)\rightarrow0$. In this semi-classical region, non-perturbative
effects are strongly suppressed and the perturbative definition of the
theory is arbitrarily good. That is, the instanton correction sum to
the left in Fig.2 gives a negligible contribution as compared to the
logarithmic one-loop correction (further higher loop corrections are
forbidden by the $N=2$ supersymmetry). Furthermore, the solitonic
magnetic monopoles that exist in the theory become very heavy and
effectively decouple.

\begin{figure} \epsfysize=3.0in \centerline{\epsffile{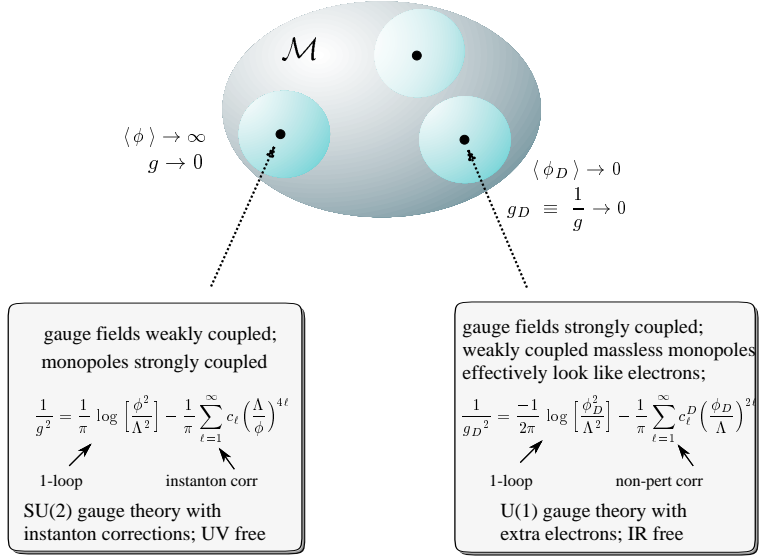}}
\caption{The exact quantum moduli space of $N=2$ supersymmetric
$SU(2)$ Yang-Mills theory has one singularity at weak coupling and two
singularities in the strong coupling region. The latter are caused by
magnetic monopoles becoming massless for the corresponding VEVs of the
Higgs field. One can go between the various regions by analytic
continuation, ie., by resumming the non-perturbative instanton
corrections in terms of suitable variables. \hfill \label{fig:SW}}
\end{figure}

However, when we now start moving in the moduli space $\cal M$ away
from the weak coupling region, the non-perturbative instanton sum to
the left in Fig.2 will less and less well converge, and the original
perturbative definition of the theory will become worse and worse.
When we are finally close to one of the other two singularities in
Fig.2, the original perturbative definition is blurred out and does
not make sense any more. The problem is thus how to suitably {\em
analytically continue} the complex gauge coupling outside the region
of convergence of the instanton series. The way to do this is to
resum the instanton series in terms of another variable, $\phi_D$, to
yield another expression for the gauge coupling that is well defined
near $1/g\rightarrow0$. That is, there is a ``dual'' Higgs field,
$\phi_D$, in terms of which the dual gauge coupling, $g_D(\phi_D)$,
makes sense in the strong coupling region of the parameter space
$\cal M$. Indeed, $\phi_D$ becomes small in this region, so that the
infinite series for the dual coupling $g_D$ to the right in Fig.2
converges very well.

The important point to realize here is that the perturbative physics
in the strong coupling region is completely different as compared to
the perturbative physics in the weak coupling region that we started
with~! At weak coupling, we had a non-abelian $SU(2)$ gauge theory,
while at strong coupling we have now an abelian $U(1)$ gauge theory
plus some extra massless matter fields (``electrons''). But the
latter only manifest themselves as elementary fields if we express
the theory in terms of the appropriate dual variables; in the
original variables, these ``electrons'' that become light at strong
coupling, are nothing but some of the solitonic magnetic monopoles
that are very massive in the weak coupling region.

All this said, we still do not know how to actually solve the theory
and determine all the unknown non-perturbative instanton coefficients
$c_\ell$, $c^D_\ell$ in Fig.2. A direct computation would be beyond
what is currently possible with ordinary field theory methods. The
insight of Seiberg and Witten\cite{SW}\ was to realize that the {\it
patching together of the known perturbative data in a globally
consistent way} is so restrictive that it fixes the theory, and
ultimately gives explicit numbers for the instanton coefficients
$c_\ell$ and $c^D_\ell$. This shows that sometimes much can be gained
by not only looking to a theory at some given fixed background VEV, but
rather by looking to a whole family of vacua, i.e., to global
properties of the moduli space.

\subsection{The message we can abstract from the field theory example}

\begin{figure}
\epsfysize=1.2in
\centerline{\epsffile{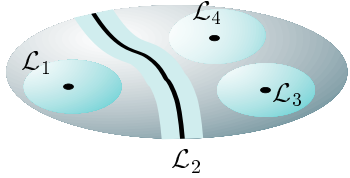}}
\caption{{The moduli space of a generic supersymmetric theory is
covered by coordinate patches, at the center of each of which the
theory is weakly coupled when choosing suitable local variables. A
local effective lagrangian exists in each patch, representing a
particular perturbative approximation. None of such lagrangians is
more fundamental than the other ones. }
\hfill
\label{fig:global}}
\end{figure}

The lesson is that one and the same physical theory can have many
perturbative descriptions. These can look completely different, and
can involve different gauge groups and matter fields.
There is in general no absolute notion of what would be weak
or strong coupling; rather what we call weak coupling or strong
coupling, or an elementary or a solitonic field, depends to some
extent on the specific description that we use. Which description
is most suitable, and which physical degrees of freedom will be light
or weakly coupled (if any at all), depends on the region of the
parameter space we are looking at.

More mathematically speaking, an effective lagrangian description makes
sense only in {\em local coordinate patches} covering the parameter
space $\cal M$ -- see Fig.3. These describe different perturbative
approximations of the same theory in terms of different weakly coupled
physical local degrees of freedom (eg, electrons or monopoles). No
particular effective lagrangian is more fundamental than any other one.
In the same way that a topologically non-trivial manifold cannot be
covered by just one set of coordinates, there is in general no globally
valid description of a family of physical theories in terms of a single
lagrangian.

It is these ideas that carry over, in a refined manner, to string
theory and thus to grand unification; in particular, they have us
rethink concepts like ``distinguished fundamental degrees of
freedom''. String moduli spaces will however be much more complex
than those of field theories, due to the larger variety of possible
non-perturbative states.

\subsection{$P$-branes and non-perturbative states in string theory}

String compactifications on manifolds $X$ are not only complex
because of the large moduli spaces they generically have, but also
because the spectrum of physical states becomes vastly more
complicated. In fact, when going down in the dimension by
compactification, there is a dramatic {\em proliferation of
non-perturbative states}.

The reason is that string theory is not simply a theory of strings:
there exist also higher dimensional extended objects, so called
``$p$-branes'', which have $p$ space and one time dimensions (in this
language, strings are 1-branes, membranes 2-branes etc; generically,
$p=0,1,...,9$). Besides historical reasons, string theory is only
called so because strings are typically the lightest of such extended
objects. In the light of duality, as discussed in the previous
sections, we know that there is no absolute distinction between
elementary or solitonic objects. We thus expect $p$-branes to play a
more important r\^ole than originally thought, and not necessarily to
just represent very heavy objects that decouple at low energies.

\begin{figure}
\epsfysize=2.0in
\centerline{\epsffile{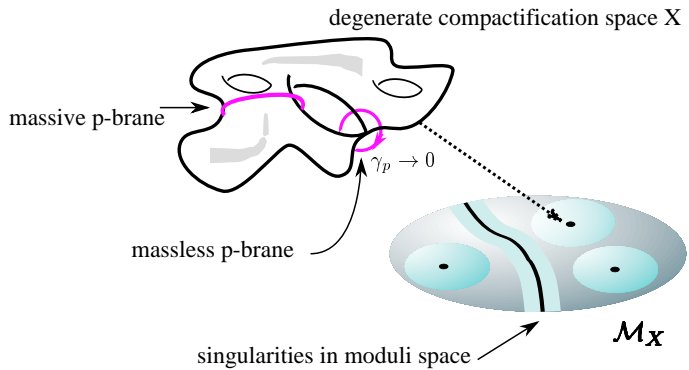}}
\caption{Non-perturbative particle-like states arise from wrapping of
$p$-branes around $p$-dimensional cycles $\gamma_p$ of the
compactification manifold $X$. These are generically very massive, but
can become massless in regions of the moduli space where the $p$-cycles
shrink to zero size. The singularities in the moduli space
are the analogs of the monopole singularities in Fig.2.
\hfill
\label{fig:branewrap}}
\end{figure}

More specifically, such $(p+1)$ dimensional objects can wrap around
$p$-dimensional cycles ${\gamma_p}$ of a compactification manifold, to
effectively become particle-like excitations in the lower dimensional
(say, four dimensional) theory. These extra solitonic states are
analogs of the magnetic monopoles that had played an important r\^ole
in the $N=2$ supersymmetric Yang-Mills theory. Since such monopoles can
be light and even be the dominant degrees of freedom for certain
parameter values, we expect something similar for the wrapped
$p$-branes in string compactifications. In fact, the volumina of
$p$-dimensional cycles ${\gamma_p}$ of $X_n$ depend on the deformation
parameters, and there are singular regions in the moduli space where
such cycles shrink to zero size (``vanishing cycles'') -- see Fig.4.
Concretely, if a $p$-dimensional cycle ${\gamma_p}$ collapses, then a
$p$-brane wrapped around ${\gamma_p}$ will give a massless state in
$D=10-n$ dimensional space-time.\cite{trans} This is because the mass
formula for the wrapped brane involves an integration over $\gamma_p$:
$$ {m_{p-{\rm
brane}}}^2\ =\  |\!\int_{\gamma_p}\Omega\,|^2\ \longrightarrow\
0\qquad\ {\rm if}\ \gamma_p\rightarrow0\ .
$$

The larger the dimension of the compactification manifold $X_n$ is (and
the lower the space-time dimension $D=10-n$), the more complicated the
topology of it will be, ie., the larger the number of ``holes'' around
which the branes can wrap. For a six dimensional Calabi-Yau space
$X_6$, there will be generically an abundance of extra non-perturbative
states that are not seen in traditional string perturbation theory.
These can show up in $D=4$ as ordinary gauge or matter fields. It is
the presence of such non-perturbative, potentially massless states that
is the basis for many non-trivial dual equivalences of string theories.

When going back to ten dimensions by making the compactification
manifold very large, these states become arbitrarily heavy and
eventually decouple. In this sense, a string model has after
compactification many more states that were not present in ten
dimensions before. The non-perturbative spectrum is very finely tuned:
in an analogous way that taking out a perturbative string state from
the spectrum would destroy modular invariance (which is a global
property of the 2d world-sheet) and thus would ruin perturbative
consistency, taking out a non-perturbative state would destroy duality
symmetries (which are a global property of the compactification
manifolds): it would make the global behavior of the theory over the
moduli space inconsistent.

\subsection{Stringy geometry}

We have seen in section 2.1 that $N=2$ supersymmetric Yang-Mills theory
is sometimes better described in terms of dual ``magnetic'' variables,
namely when we are in a region of the moduli space where certain
magnetic monopoles are light. In this dual formulation, the originally
solitonic monopoles look like ordinary elementary, perturbative degrees
of freedom (``electrons''). Analogously, dual formulations exist also
for string theories, in which non-perturbative solitons are described
in terms of weakly coupled ``elementary'' degrees of freedom. It was
one of the breakthroughs in the field when it was realized how this
exactly works: the relevant objects dual to certain solitonic states
are special kinds of ($p$-)branes, so-called
``$D(p$-)branes''.\cite{Dbr} Due to the many types of $p$-branes on the
one hand, and the large variety of possible singular geometries of the
manifolds $X$ on the other, the general situation is however very
complicated. It is easiest to describe it in terms of typical examples.

For instance, massless or almost massless non-abelian gauge bosons
$W^\pm$ belonging to $SU(2)$ can be equivalently described in a
number of dual ways (see Fig.5):

\begin{figure}
\epsfysize=1.4in
\centerline{\epsffile{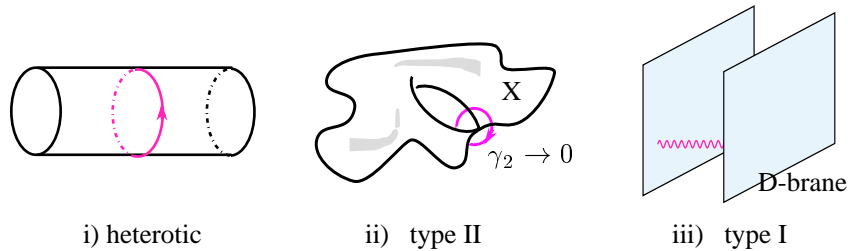}}
\caption{Different geometries can parametrize one and the same
physical situation. Shown here are three dual ways to represent
an $SU(2)$ gauge field and the associated Higgs mechanism.
\hfill \label{fig:gaugemech}} \end{figure}

\noindent {\bf i)} In the heterotic string compactified on some
higher dimensional torus, a massless gauge boson is represented by a
fundamental heterotic string wrapped around part of the torus, with a
certain radius (say $R=1$ in some units; changing the orientation of
the string maps $W^+\leftrightarrow W^-$).This is a perturbative
description, since it involves an elementary string. If the radius
deviates from $R=1$, the gauge field gets a mass, providing a
geometrical realization of the Higgs mechanism.

\noindent {\bf ii)} In the compactified type IIA string, the gauge
boson arises from wrapping a 2-brane around a collapsed 2-cycle
$\gamma_2$ of $X$. This is a non-perturbative formulation in terms of
string theory. If $\gamma_2$ does not quite vanish, the gauge field
retains some non-zero mass, thereby realizing the Higgs mechanism in
a different manner.

\noindent {\bf iii)} In the type I string model, an $SU(2)$ gauge
boson is realized by an open string stretched between two flat
$D$-branes. This is another perturbative formulation of the Higgs
mechanism. The mass of the gauge boson is proportional to the length
of the open string, and thus vanishes if the two $D$-branes move on
top of each other.

We thus see that very different mathematical geometries can represent
the identical physical theory, here the $SU(2)$ Higgs model -- these
geometries really should be identified in string theory. This provides
a special example of a more general concept, which is about getting
better and better understood: ``stringy geometry''. In stringy
geometry, certain mathematically different geometries are treated as
equivalent, and just seen as different choices of ``coordinates'' in
some abstract space of string theories. The underlying physical idea is
that while in ordinary geometry point particles are used to measure
properties of space-time, in stringy geometry one augments this by
string and other $p$-brane probes. It is the wrappings and stretchings
that these extra objects are able to do that can wash out the
difference between topologically and geometrically distinct manifolds.

In effect, a string theory of some type ``$A$'' when compactified on
some manifold $X_A$, can be dual to another string theory ``$B$'' on
some manifold $X_B$ -- and this even if $A$, $B$ and/or $X_A$ and $X_B$
are profoundly different.\cite{HT} Again, all what matters is that the
full non-perturbative theories coincide, while there is no need for the
perturbative approximations to be even remotely similar.

\section{The Grand Picture}

\subsection{Dualities of higher dimensional string theories}

We are now prepared to go back to ten dimensions and revisit the five
perturbatively defined string models of section 1.2. In view of the
remarks of the preceding sections concerning the irrelevance of
perturbative concepts, we will now find it perhaps not too surprising
to note that these five theories are really nothing but different
approximations of just {\rm one} theory. In complete analogy to what
we said about $N=2$ supersymmetric gauge theory in section 2.1, they
simply correspond to certain choices of preferred ``coordinates''
that are adapted to specific parameter regions.

Although these facts can be stated in such simple terms, they are
so non-trivial that it took more than a decade to discover them. It
can now be explicitly shown that by compactifying any one of
the five theories on a suitable manifold, and then de-compactifying it
in another manner, one can reach any other of the five theories in a
continuous way.

\begin{figure}
\epsfysize=1.2in
\centerline{\epsffile{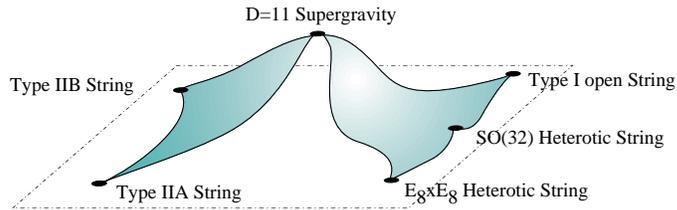}}
\caption{Moduli space in ten and eleven dimensions. Its
singular asymptotic regions correspond to the five well-known
supersymmetric string theories in $D=10$, plus an eleven dimensional
$M$-theory. The vertical direction roughly reflects the
space-time dimension.
\hfill
\label{fig:bigPict10d}}
\end{figure}

 A surprise happened when the strong coupling limit of the type
IIA string was investigated\cite{Weleven}: it turned out that in this
limit, certain non-perturbative ``$D0$-branes'' form a continuous
spectrum and effectively generate an extra 11th dimension. That is,
at ultra strong coupling the ten-dimensional type IIA string theory
miraculously gains 11-dimensional Lorentz invariance, and the
low-energy theory turns into $D=11$ supergravity. This was especially
surprising because eleven dimensional supergravity is not a low
energy limit of a string theory, but rather seems to be related to
supermem\-branes. In other words, non-perturbative dualities take us
beyond string theory~!

So what we have are not five but six $10$- or $11$-dimensional
approximations, or local coordinate patches on some moduli space -- see
Fig.6. But to what entity are these theories approximations~? Do we
have here the moduli space of some underlying microscopic theory~?
Indeed there is a candidate for such a theory, dubbed
``$M$-theory''\cite{BFSS,Mtheory}. Its low energy limit does give
$D=11$ supergravity, and simple compactifications of it give all of the
five string theories in $D=10$. It may well be that $M$-theory,
currently under intense investigation, holds the key for a detailed
understanding of non-perturbative string theory. However, since
$M$-theory is (presumably) a main topic of Susskind's lecture in these
proceedings, we will not discuss any further details here. We will
rather return to the lower dimensional theories.

\subsection{Quantum moduli space of four-dimensional strings}

Fig.6.\ shows only a small piece of a much more extended moduli space,
namely only the piece that describes higher dimensional theories.
These are relatively simple and there is only a small number of them.
As discussed above, the lower dimensional theories obtained by
compactification are much more intricate.

Among the best-investigated string theories in four dimensions are
the ones with $N=2$ supersymmetry -- these are the closest analogs of
the $N=2$ gauge theory that is so well understood. They can be
obtained by compactifications of type IIA/B strings on Calabi-Yau
manifolds $X_6$, or dually, by compactifications of heterotic strings
on\footnote{Here, $T_2$ denotes the two-torus, and $K3$ the four
dimensional version of a Calabi-Yau space.} $K3\times T_2$. Since
there are roughly 10$^4$ of such Calabi-Yau manifolds $X_6$, the
complete $D=4$ string moduli space will have roughly 10$^4$
components, each with typical dimension 100 (keeping in mind that
there can be non-trivial identifications between parts of this moduli
space)-- see Fig.7. We see that the moduli space is drastically more
complicated as it is either for the high-dimensional theories, or for
the $N=2$ gauge theory. Each of the 10$^4$ components typically has
several ten or eleven dimensional decompactification limits, so that
one should imagine very many connections between the upper and lower
parts of Fig.7 (indicated by dashed lines).

\begin{figure} \epsfysize=3.0in \centerline{\epsffile{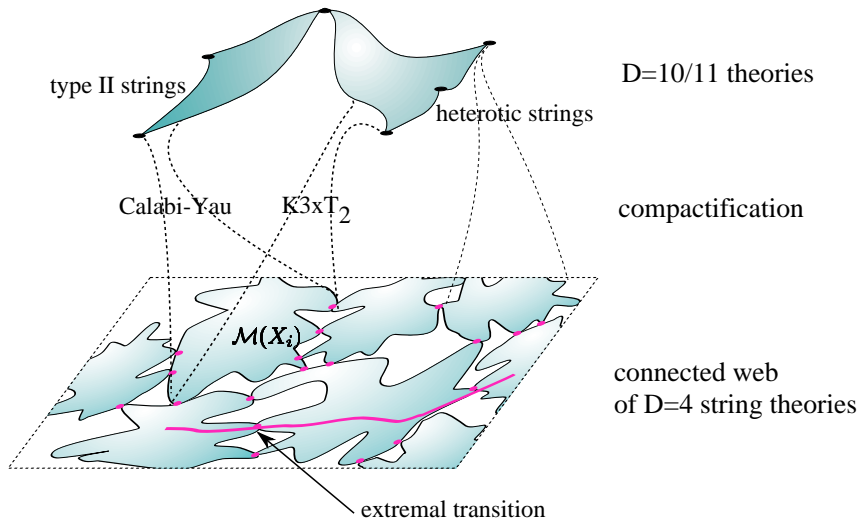}}
\caption{Rough sketch of the space of $N=2$ supersymmetric string
theories in four dimensions. In general, a given region in the 4d
moduli space can be reached in several dual ways via compactification
of the higher dimensional theories. Each of the perhaps 10$^4$
components corresponds typically to a 100 dimensional family of
perturbative string vacua, and represents the moduli space of a single
Calabi-Yau manifold (like the one shown in Fig.4). Non-perturbatively,
all these vacua turn out to be connected by extremal transitions and
thus form a continuous web. \hfill \label{fig:bigPict4d}} \end{figure}

An interesting fact is that all of the known\footnote{Strictly
speaking, we cannot exclude further, disconnected components to exist.}
$10^4$ families of perturbative $D=4$ string vacua are connected by
non-perturbative extremal transitions. To understand what we mean by
that, simply follow a path in the moduli space as indicated in Fig.7,
starting somewhere in the interior of a blob. The massive spectrum will
continuously change, and when we hit singularities, extra massless
states appear and perturbation theory breaks down.

It can in particular happen that a non-perturbative massless
Higgs field appears, to which we can subsequently give a vacuum
expectation value to deform the theory in a direction that was not
visible before. In this way, we can leave the moduli space of a single
perturbative string compactification, and enter the moduli space of
another one. Thus, non-perturbative extra states can glue together
different perturbative families of vacua in a physically smooth
way.\cite{trans} It can be proven that one can connect in this
manner all of the roughly 10$^4$ known components that were previously
attributed to different four dimensional string theories. In other
words, the full non-perturbative quantum moduli space of $N=2$
supersymmetric strings seems to form {\em one} single entity.

This is not much of a practical help for solving the vacuum degeneracy
problem, but it is conceptually satisfying: instead having to choose
between many four dimensional string theories, each one equipped with
its own moduli space, we really have just one theory with however very
many facets.

This as far as $N=2$ supersymmetric strings in four dimensions are
concerned -- the situation will still be much more complicated for the
phenomenologically important $N=1$ supersymmetric string theories,
whose investigation is currently under way. The main novel features
that can be addressed in these theories are chirality and
supersymmetry breaking. It seems that certain aspects of these
theories are best described by choosing still another dual
formulation, called ``$F$-theory''\cite{Ftheory}. This is a
construction formally living in twelve dimensions, and whether it is
simply a trick to describe certain features in an elegant fashion, or
whether it is a honest novel theory in its own, remains to be
seen.

\section{``Theoretical experiments''}

How can we convince ourselves that the considerations of the previous
sections really make sense~? Clearly, all what we can do for the
foreseeable future to test these ideas are consistency checks. Such
checks can be highly non-trivial, from a formal as well as
from a physical point of view. So far, numerous qualitative and
quantitative tests have been performed, and not a single test on the
dualities has ever failed~!

\noindent
To give some flavor, let us recapitulate a few characteristic checks:

\goodbreak

 \noindent {\bf i)} {\em Complicated perturbative corrections} can be
computed and compared for string models that are dual to each
other. In all cases, the results perfectly agree. Consider for example
certain 3-point couplings $\kappa$ in $N=2$ supersymmetric string
compactifications. As mentioned before, such theories can be obtained
either from type II strings on Calabi-Yau manifolds $X_6$,
or dually from heterotic string compactifications on $K3\times T_2$. In
the type II formulation, these couplings can be computed (via ``mirror
symmetry''\cite{mirror}) by counting world-sheet instantons (embedded
spheres) inside the Calabi-Yau space. Concretely, in a specific model
the result takes the following explicit form:
\def\eft{E_4(T)}\def\est{E_6(T)}\def\efu{E_4(U)}\def\esu{E_6(U)}
$$
\kappa\ =\ {i\over 2\pi}
{\eft\efu\esu(\eft^3-\est^2)\over \efu^3\est^2-\eft^3\esu^2}\ ,
$$
in terms of certain modular functions $E_4$, $E_6$ depending on moduli
fields $T,U$. The very same expression can be obtained also in a
completely different manner, namely by performing a standard one-loop
computation in the dual heterotic string model -- in precise agreement
with the postulated string duality; see Fig.8. Many similar tests, also
involving higher loops and gravitational couplings, have been shown to
work out as well.

\begin{figure}
\epsfysize=1.0in
\centerline{\epsffile{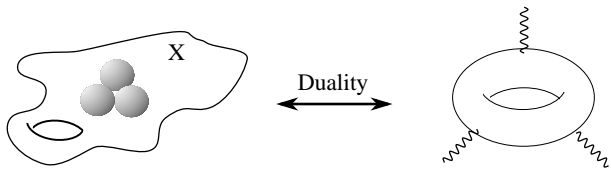}}
\caption{Counting spheres in a Calabi-Yau space $X$ in type IIA
string theory, leads ultimately to the same expressions for certain
low energy couplings as a standard one-loop computation in the
heterotic string.
\hfill
\label{fig:mirrorsym}}
\end{figure}

\noindent {\bf ii)}
{\em State count in black holes.}
This is a highly non-trivial physics test. The issue is to compute the
Bekenstein-Hawking entropy $S_{BH}$ (=area of horizon) of an
extremal (or near-extremal) black hole. Strominger and Vafa\cite{SV}\
considered the particular case of an extremal $N=4$ supersymmetric
black hole in $D=5$, where one knows that
$$
S_{BH}\ =\ 2\pi\sqrt{{q_fq_h\over2}}\ ,
$$
where $q_f$, $q_h$ are the electric and axionic charges of the black
hole. The idea is to use string duality to represent a large
semi-classical black hole in terms of a type IIB string
compactification on $K3\times S^1$. This eventually boils down to a
2d sigma model on the moduli space of a gas of $D0$-branes on $K3$.
Counting states in this model indeed exactly reproduces the above
entropy formula for large charges.


This test (and refinements of it\cite{BHreviews})
does not only add credibility to
the string duality claims from a new perspective, but also tells that
there are {\em no missing degrees of freedom} that we might have been
overlooking -- any theory of quantum gravity better comes up with the
same count of relevant microscopic states.

\noindent {\bf ii)} {\em Recovering exact non-perturbative field theory
results.} One can derive the exact solution of the $N=2$ supersymmetric
gauge theory described in section 2.1 as a consequence of string
duality. The point is that duality often maps classical into quantum
effects and vice versa. This makes it for example
possible\cite{qmirror}\ to obtain with a {\em classical} computation in
the compactified type II string, certain non-perturbative results for
the compactified heterotic string. In particular, counting world-sheet
instantons similarly as in point i) above, and suitably decoupling
gravity (see Fig.9), allows to exactly reproduce\cite{KKLMV}\ the
non-perturbative effective gauge coupling $g(\phi)$ of Fig.2.

\begin{figure}
\epsfysize=1.8in
\centerline{\epsffile{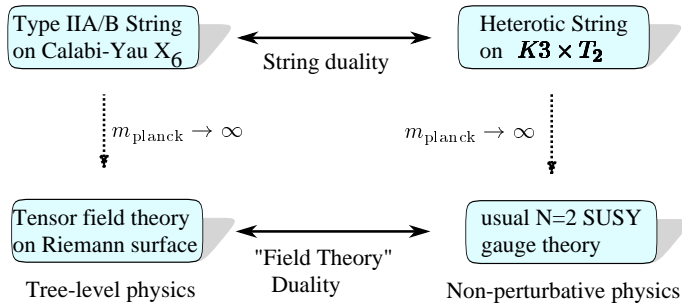}}
\caption{Recovering Seiberg-Witten theory from string duality. In the
field theory limit, one sends the Planck mass to infinity in order to
decouple gravity and other effects that are not important. A
remnant of the string duality survives, which is a
duality between the standard formulation of the gauge
theory, and a novel one that was not known before. It gives a physical
interpretation of the geometry underlying the non-perturbative
solution of the $N=2$ gauge theory.
\hfill
\label{fig:rigid}}
\end{figure}

This is very satisfying, as it gives support to both the underlying
string duality and the solution of the $N=2$ gauge theory. Indeed,
while the basic heterotic-type II string duality\cite{HT}\ and the
Seiberg-Witten theory\cite{SW}\ have been found independently from each
other around the same time, their mutual compatibility was shown only
later.

It is the clearly visible, convergent evolution of a priori separate
physical concepts, besides overwhelming ``experimental'' evidence, what
gives the string theorists confidence in the validity of their ideas.

\section{Spinoff: $D$-brane technology,
  and novel quantum theories without gravity.}

The techniques for obtaining exact non-perturbative results for
ordinary field theories from string duality are not limited to only
reproducing results that one already knows -- in fact, they have opened
the door for deriving new results for a whole variety of
field theories in various dimensions.

More specifically, there are currently two main approaches (related by
duality) for obtaining standard and non-standard field theories from
string theory -- each one has its own virtues. One can either study the
singular geometry of Calabi-Yau or other compactification manifolds,
and consider the effect of wrapped branes
-- much like it is depicted in Fig.5 part ii).

Alternatively, one can model\cite{HW}\ the relevant string geometry
in terms of parallel $D$-branes, with open strings and other branes
running between them (remember that $D$-branes are
roughly perturbative duals of $p$-brane solitons). This refers to
part iii) of Fig.5. For example, $N=2$ Yang-Mills theory in four
dimensions can be represented by a configuration of $D$-branes as
shown in Fig.10 a). From this simple picture one can reproduce
the non-perturbative solution of the gauge theory\cite{SWfromM}.
Similar arrangements can describe $N=1$ supersymmetric gauge theories
as well, like supersymmetric QCD (Fig.10 b)). Some qualitative
features, like chiral symmetry breaking or confinement, can be seen
in this model.\cite{MQCD}

\begin{figure}
\epsfysize=1.7in
\centerline{\epsffile{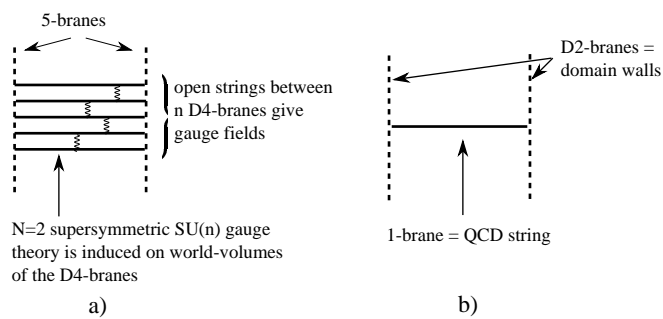}}
\caption{$D$-brane technology allows to represent gauge theories
in a dual way, in which non-trivial properties are
encoded in simple geometric pictures.
\hfill
\label{fig:Dbranes}}
\end{figure}

In addition, somewhat unexpectedly,  these methods have also led to
the discovery of completely new kinds of quantum theories that were not
known to exist before. More specifically, when decoupling gravity there
is no reason why one should always end up with a standard field theory
that one already knows, like gauge theory coupled to some matter
fields. Rather, by either looking to specific singular geometries or to
appropriate $D$-brane configurations, one has found a number of exotic
limits.\cite{Weleven,sixdim} Such theories are typically strongly
coupled and do not have any known description in terms of traditional
quantum field theory; they comprise (non-gravitational) tensionless
strings and novel non-trivial IR fixed points. The main indication that
they really exist is that they arise as decoupling limits of a larger
string (or $M$-) theory that is consistent by itself.

Especially interesting in this context are exotic theories that
serve as transition points between $N=1$ string vacua with different
net numbers of chiral fermions. Such smooth chirality changing
processes are not possible in conventional field theory, but they can
occur string theory.\cite{chiral} This opens up the possibility that
$N=1$ supersymmetric string vacua are (to some extent)
unified in a manner analogous to the $N=2$ vacua,
similar to but much more complicated as hinted at in Fig.7.


\lref\SWreviews{For reviews see, for example:\\
{L.\ Alvarez-Gaum\'e and S.\ Hassan,
 \nihil{Introduction to S-Duality in N=2
 supersymmetric Gauge Theories:
A pedagogical review of the Work of Seiberg and Witten,}
 Fortsch.\  Phys.\ {\bf 45} (1997) 159-236,
 \eprt{hep-th/9701069};}\\
{W.\ Lerche, \nihil{Introduction to Seiberg-Witten Theory
and its Stringy Origin},
Nucl.\ Phys.\ Proc.\ Suppl.\ {\em 55B} (1996) 83,
\eprt{hep-th/9611190};\\
{M.\ Peskin,
 \nihil{Duality in Supersymmetric Yang-Mills Theory,}
lectures at TASI-96 Summer School, \eprt{hep-th/9702094}.}
}
}

\lref\SW{N.\ Seiberg and E.\ Witten, \nihil{Electric - magnetic
duality, monopole condensation, and confinement in N=2 supersymmetric
Yang-Mills theory,} \nup426(1994) 19, \eprt{hep-th/9407087};
\nihil{Monopoles, duality and chiral symmetry breaking in N=2
supersymmetric QCD,} \nup431(1994) 484, \eprt{hep-th/9408099}.}

\lref\intro{For introductory textbooks, see e.g.:\\
M.\ Green, J.\ Schwarz and E.\ Witten, {\em Superstring Theory},
Vol.\ 1-2, Cambridge University Press 1987;\\
D.\ L\"ust and S.\ Theisen, {\em Lectures on String Theory},
Lecture Notes in Physics 346, Springer 1989.\\
For more recent reviews, see e.g.:\\
{J.\ Polchinski,
 \nihil{What is string theory?,}
 lectures presented at the 1994 Les Houches Summer School,
 \eprt{hep-th/9411028};} {
 \nihil{String duality: A Colloquium,}
 Rev.\  Mod.\  Phys.\ {\bf 68} (1996) 1245-1258,
 \eprt{hep-th/9607050};}\\
{K.\ Dienes,
 \nihil{String theory and the path to unification: A Review of recent
developments,}
 Phys.~ Rept.~{\bf 287} (1997) 447-525,
 \eprt{hep-th/9602045};}\\
{C.\ Vafa,
 \nihil{Lectures on strings and dualities,}
lectures at ICTP Summer School 1996,
 \eprt{hep-th/9702201};}\\
{S.\ Mukhi,
 \nihil{Recent developments in string theory: A Brief review for
particle physicists,}
 \eprt{hep-ph/9710470}.}
}

\lref\KD{K.\ Dienes, as cited in ref.\ \cite{intro}.}

\lref\brian{For a review, see:
{B.\ Greene,
 \nihil{String theory on Calabi-Yau manifolds,}
lectures at TASI-96 Summer School, \eprt{hep-th/9702155}.}
}

\lref\Dbr{J.\ Polchinski,
 \nihil{Dirichlet-Branes and Ramond-Ramond Charges,}
 Phys.\  Rev.\  Lett.\ {\bf 75} (1995) 4724-4727,
 \eprt{hep-th/9510017}; \\
 for a review, see:
{J.\ Polchinski,
 \nihil{Tasi Lectures on D-Branes,} TASI-96 Summer School,
 \eprt{hep-th/9611050}.}}

\lref\BFSS{T.\ Banks, W.\ Fischler, S.\ Shenker and L.\ Susskind,
 \nihil{M-Theory as a Matrix Model: A Conjecture,}
 Phys.\  Rev.\ {\bf D55} (1997) 5112-5128,
 \eprt{hep-th/9610043}.}

\lref\Mtheory{For reviews, see e.g.: \\
{A.\ Bilal, \nihil{M(atrix) Theory: a Pedagogical Introduction,}
\eprt{hep-th/9710136.};}\\
T.\ Banks, \nihil{Matrix Theory,} Trieste Spring School 1997,
 \eprt{hep-th/9710231}.
}

\lref\Weleven{E.\ Witten,
 \nihil{Some Comments on String Dynamics,} \eprt{hep-th/9507121}.}

\lref\sixdim{See for example:
{N.\ Seiberg, E.\ Witten,
 \nihil{Comments on string dynamics in six-dimensions,}
 Nucl.~ Phys.~{\bf B471} (1996) 121-134,
 \eprt{hep-th/9603003};}\\
{O.\ Aharony, M.\ Berkooz, S.\ Kachru, N.\ Seiberg and E.\ Silverstein,
 \nihil{Matrix description of interacting theories in six dimensions,}
 \eprt{hep-th/9707079}.}
}

\lref\HT{C.\ Hull and
P.\ Townsend, \nihil{Unity of superstring dualities,}
\nup438 (1995) 109, \eprt{hep-th/9410167}.}

\lref\gaugerev
{For a review on supersymmetric gauge theories, see:\\
K.\ Intriligator and N.\ Seiberg,
 \nihil{Lectures on supersymmetric gauge theories and
  electric - magnetic duality,}
 Nucl.~ Phys.~ Proc.~ Suppl.~{\bf 45BC} (1996) 1-28,
 \eprt{hep-th/9509066}.}

\lref\Ftheory{{C.\ Vafa,
 \nihil{Evidence for F-theory,}
 Nucl.\ Phys.\ {\bf B469} (1996) 403-418,
 \eprt{hep-th/9602022}.}}

\lref\qmirror{
{S.\ Kachru and C.\ Vafa,
 \nihil{Exact Results for N=2 Compactifications of Heterotic Strings,}
 Nucl.\  Phys.\ {\bf B450} (1995) 69-89,
 \eprt{hep-th/9505105};}\\
{S.\ Ferrara, J.\ Harvey, A.\ Strominger and C.\ Vafa,
 \nihil{Second quantized Mirror Symmetry,}
 Phys.\  Lett.\ {\bf B361} (1995) 59-65,
 \eprt{hep-th/9505162}.}}

\lref\trans{A. Strominger,
 \nihil{Massless black holes and conifolds in string theory,}
 \nup451 (1995) 96,
\eprt{hep-th/9504090};\\ {for a review, see: A. Strominger,
\nihil{Black Hole Condensation and Duality in String
Theory,} \eprt{hep-th/9510207}.}}

\lref\BHreviews{For reviews see:
{J.\ Maldacena,
 \nihil{Black Holes in String Theory,}
PhD thesis, \eprt{hep-th/9607235}; {
 \nihil{Black Holes and D-Branes,}
 \eprt{hep-th/9705078}.}
}}

\lref\SV{A.\ Strominger and C.\ Vafa,{
 \nihil{Microscopic origin of the Bekenstein-Hawking entropy,}
 Phys.\  Lett.\ {\bf B379} (1996) 99-104,
 \eprt{hep-th/9601029}.}
}

\lref\SWfromM{E.\ Witten,
 \nihil{Solutions of four-dimensional Field Theories via M-Theory,}
 \eprt{hep-th/9703166}.}
\lref\MQCD{E.\ Witten,
 \nihil{Branes and the Dynamics of QCD,}
 \eprt{hep-th/9706109}.}

\lref\GM{G.\ Moore,
 \nihil{Atkin-Lehner symmetry,}
 Nucl.\ Phys.\ {\bf B293} (1987) 139.}

\lref\cosmo{E.\ Witten,
 \nihil{Strong coupling and the cosmological constant,}
 Mod.~ Phys.~ Lett.~{\bf A10} (1995) 2153-2156,
 \eprt{hep-th/9506101}.}

\lref\mirror{S.\ Yau (ed.), {\it Essays on Mirror Manifolds},
International Press 1992; B.\ Greene (ed.), {\it Mirror symmetry II},
AMS/IP Studies in advanced mathematics vol 1, 1997.}

\lref\geometric{
{A.\ Klemm, W.\ Lerche, P.\ Mayr, C.\ Vafa and
N.\ Warner,
 \nihil{Self-dual strings and N=2 supersymmetric field theory,}
 Nucl.\  Phys.\ {\bf B477} (1996) 746-766,
 \eprt{hep-th/9604034};}//
{M.\ Bershadsky at al.,
 \nihil{Geometric singularities and enhanced gauge symmetries,}
 Nucl.\  Phys.\ {\bf B481} (1996) 215-252,
 \eprt{hep-th/9605200};}//
S.\  Katz and C.\ Vafa,
 \nihil{Geometric engineering of N=1 quantum field theories,}
Nucl.\  Phys.\ {\bf B497} (1997) 196-204,
 \eprt{hep-th/9611090}.
}

\lref\HW{A.\ Hanany and E.\ Witten, \nihil{Type IIB Superstrings, BPS
Monopoles, and three-dimensional Gauge Dynamics,} Nucl.\  Phys.\ {\bf
B492} (1997) 152-190, \eprt{hep-th/9611230}.}

\lref\KKLMV{S.\  Kachru, A.\  Klemm, W.\ Lerche, P.\  Mayr and
C.\ Vafa,
 \nihil{Nonperturbative Results on the Point Particle Limit of N=2
Heterotic String Compactifications,}
 Nucl.\ Phys.\ {\bf B459} (1996) 537-558,
 \eprt{hep-th/9508155}.}

\lref\chiral{For reviews, see e.g.:
{S.\ Kachru,
 \nihil{Aspects of N=1 string dynamics,}
 \eprt{hep-th/9705173};}\\
{E.\ Silverstein,
 \nihil{Closing the Generation Gap,} Talk at Strings '97,
 \eprt{hep-th/9709209}.}
}

\section*{References}

\newcommand\nil[1]{{}}
\newcommand\nihil[1]{{\sl #1}}
\newcommand\ex[1]{}
\newcommand{\nup}[3]{{\em Nucl.\ Phys.}\ {B#1#2#3}\ }
\newcommand{\plt}[3]{{\em Phys.\ Lett.}\ {B#1#2#3}\ }
\newcommand{\prl}{{\em Phys.\ Rev.\ Lett.}\ }
\newcommand{\cmp}[3]{{\em Comm.\ Math.\ Phys.}\ {A#1#2#3}\ }

\end{document}